\documentclass[final,12pt]{elsarticle}

\makeatletter
\def\ps@pprintTitle{%
 \let\@oddhead\@empty
 \let\@evenhead\@empty
 \def\@oddfoot{\small{\it{DRAFT - not for citation without permission of the authors \hfill \today}}}%
 \let\@evenfoot\@oddfoot}
\makeatother

\usepackage{graphicx}
\usepackage{amssymb}
\usepackage{lineno}
\usepackage{hyperref}
\usepackage{array}
\usepackage{xcolor}
\usepackage{soul}
\usepackage{multirow}
\usepackage{tabularx}
\usepackage{booktabs}
\usepackage{dblfloatfix}
\usepackage{lscape}
\usepackage{rotating}
\usepackage{gensymb}

\usepackage[symbol]{footmisc}

\begin{document}

\begin{frontmatter}
%% Title, authors and addresses

\title{Exposure Density and Neighborhood Disparities in COVID-19 Infection Risk: Using Large-scale Geolocation Data to Understand Burdens on Vulnerable Communities}

%% include affiliations in footnotes:
\author[marron]{Boyeong Hong}
\author[marron]{Bartosz J. Bonczak}
\author[stern]{Arpit Gupta}
\author[langone]{Lorna E. Thorpe}

%% Specifying corresponding author: 
%%   1. use original Elsevier template with \corref function
% \author[marron,cusp]{Constantine E. Kontokosta\corref{correspondingauthor}}

% \cortext[correspondingauthor]{Corresponding author}
% \ead{ckontokosta@nyu.edu}

%%   2. Manually specify footnote to ensure it appearance on the first page
\author[marron,cusp]{Constantine E. Kontokosta$^*$}
\footnotetext[1]{Corresponding author: ckontokosta@nyu.edu} 

\address[marron]{New York University, Marron Institute of Urban Management, 60 5th Avenue, New York, NY 10011}
\address[stern]{New York University, Stern School of Business, 44 West 4th Street, New York, NY 10012}
\address[langone]{New York University School of Medicine, Department of Population Health, New York, NY, 650 First Avenue, Fifth Floor, New York, NY 10016}
\address[cusp]{New York University, Center for Urban Science and Progress, 370 Jay Street, Brooklyn, NY 11201}

\begin{abstract}
%% Text of abstract
This study develops a new method to quantify neighborhood activity levels at high spatial and temporal resolutions and test whether, and to what extent, behavioral responses to social distancing policies vary with socioeconomic and demographic characteristics. We define exposure density ($E_x\rho$) as a measure of both the localized volume of activity in a defined area and the proportion of activity occurring in non-residential and outdoor land uses, areas that are associated with an increased risk of exposure to the virus. We utilize this approach to capture inflows/outflows of people as a result of the pandemic and changes in mobility behavior for those that remain. First, we develop a generalizable method for assessing neighborhood activity levels by land use type using smartphone geolocation data over a three-month period covering more than 12 million unique users within the Greater New York area. Second, we measure and analyze disparities in community social distancing by identifying patterns in neighborhood activity levels and characteristics before and after the stay-at-home order. Finally, we evaluate the effect of social distancing in neighborhoods on COVID-19 infection rates and outcomes associated with localized demographic, socioeconomic, and infrastructure characteristics in order to identify disparities in health outcomes related to exposure risk. Our findings provide insight into the timely evaluation of the effectiveness of social distancing for individual neighborhoods and support a more equitable allocation of resources to support vulnerable and at-risk communities. Our findings demonstrate distinct patterns of activity pre- and post-COVID across neighborhoods. The variation in exposure density has a direct and measurable impact on the risk of infection. 
\end{abstract}

\begin{keyword}
Social mobility \sep neighborhood disparity \sep COVID-19 \sep Coronavirus \sep computational modeling
%% keywords here, in the form: keyword \sep keyword

%% MSC codes here, in the form: \MSC code \sep code
%% or \MSC[2008] code \sep code (2000 is the default)

\end{keyword}
\end{frontmatter}

%%
%% Start line numbering here if you want
%%
% \linenumbers{}

%% main text
\section{Introduction}
\label{S:1}
When Wuhan, China was struck by an outbreak of a novel coronavirus (2019-nCoV or COVID-19) in late 2019, scientists were concerned about the potential for rapid spread of the virus \cite{callaway2020time}. %On March 11, 2020, the World Health Organization (WHO) officially announced that COVID-19 had reached pandemic status \cite{wang2020novel, remuzzi2020covid, cucinotta2020declares, world2020coronavirus}. 
By the end of June 2020, a total of 10 million people had been infected and 500,000 people had lost their lives in more than 200 countries \cite{world2020coronavirus,van2020using}. The COVID-19 pandemic is considered the most severe public health crisis since the 1918 Spanish flu due to the transmission and infection characteristics of the disease, with few effective treatments at the time this article was written \cite{world2020coronavirus,sen2020social,courtemanche2020did,gao2020mobile}. Given the global scale of the pandemic, a coordinated response is necessary to mitigate the spread of the virus both within and across country borders \cite{remuzzi2020covid}.

Social distancing (also referred to as physical distancing) is one of the most  effective behavioral strategies to reduce the transmission rate of COVID-19 \cite{sen2020social,courtemanche2020did,chudik2020voluntary,patrick2020social,gao2020mobile,jay2020neighborhood}. Social distancing reduces the probability of contacts between individuals who might be infected, resulting in reduced exposure risk \cite{jay2020neighborhood,adolph2020pandemic}. Governments have implemented a range of social distancing policies, including travel bans, restrictions on gatherings, school closures, non-essential business closures, and restaurant restrictions. In particularly hard-hit locations, mandatory “stay-at-home” orders have been issued to limit or avoid unnecessary close contacts outside of the home \cite{adolph2020pandemic,jay2020neighborhood,chinazzi2020effect}. 

Studies have found social distancing measures help to prevent transmission of the virus and reduce the reproduction (R\textsubscript{0}) number \cite{gao2020mobile,chudik2020voluntary,jay2020neighborhood,abouk2020immediate,hatchett2007public,farboodi2020internal,matrajt2020evaluating}. These practices help to avoid overwhelming hospital intensive care units (ICUs) and health care systems, control doubling time of infections, and ultimately save lives \cite{gao2020mobile,greenstone2020does,adolph2020pandemic}. Although not without potentially significant hardship to individuals and communities, social distancing is an important public health tool to flatten the epidemic curve and support longer-term economic and public health benefits \cite{sen2020social,greenstone2020does,thunstrom2020benefits}. 

However, the impact of, and response to, stay-at-home orders, and social distancing guidelines more broadly, is not uniform across neighborhoods and communities. In order to maximize the positive effects of social distancing, individuals need to change their typical behavior, often dramatically \cite{van2020using,sen2020social}. Despite government-mandated social distancing policies (e.g. New York State's PAUSE order), socio-behavioral responses vary across neighborhoods, further contributing to disparities in risk of infection \cite{courtemanche2020did,jay2020neighborhood}. Disparities in social distancing practices, namely geographic or population subgroup differences in adopting behavior changes in response to the same policy context, may stem from varying levels of awareness, perception, or belief in the severity of the virus threat; differences in social and cultural norms; or the ability of households and communities to alter normal activity patterns given economic constraints or other existing responsibilities \cite{van2020using, jay2020neighborhood,wise2020changes,caley2008quantifying}. For example, lower-income households typically do not have the option to work from home and going to a place of work (often in essential services) is unavoidable, meaning higher risk of exposure to COVID-19 for themselves as well as their families and communities \cite{jay2020neighborhood,atchison2020perceptions}. Within specific neighborhoods and communities, norms can also be reinforcing; if large numbers of residents are essential workers and not socially distancing, other residents may have similar behavioral responses \cite{van2020using}.

A growing number of outbreaks are occurring in densely populated areas \cite{desai2020urban}, with disproportionate impacts on low-income and minority communities. Measuring and understanding social distancing and behavior change across neighborhoods during a pandemic can provide critical insight to the design and implementation of more effective - and equitable - public health policy. Given the potential heterogeneity in localized responses to social distancing recommendations, quantifying local patterns of activity represents an emerging tool to understand and eventually reduce local exposure risk and limit community outbreaks \cite{jay2020neighborhood}. Although there has been increasing awareness of the troubling disparities in infection rates and outcomes in vulnerable communities, the effectiveness of behavioral interventions at the scale of individual neighborhoods has not been fully studied. Often, studies that do attempt to observe effects at higher spatial resolutions rely on simulations or are limited to relatively coarse areal units (e.g. county or state) due to data limitations and computational constraints. 

This study develops a new method to quantify neighborhood activity levels at high spatial and temporal resolutions. We define exposure density ($E_x\rho$) as a measure of both the localized volume of activity in a defined area and the proportion of activity occurring in non-residential and outdoor land uses, areas that are associated with an increased risk of exposure to the virus. We utilize this approach to capture inflows/outflows of people as a result of the pandemic and changes in mobility behavior for those that remain. We test whether, and to what extent, behavioral responses to social distancing policies vary with socioeconomic and demographic characteristics. We focus on New York City (NYC), the first epicenter of the pandemic in the U.S., which enacted a statewide ``stay-at-home" order (NY on PAUSE) on March 22. Through June 30, NYC had more than 212,000 confirmed cases of COVID-19, accounting for 8\% of the nationwide total, resulting in at least 18,492 confirmed deaths and 4,604 probable deaths. Our methodology proceeds in three steps. First, we develop a generalizable method for assessing neighborhood activity levels using smartphone geolocation data over a three-month period covering more than 12 million unique users within the Greater New York area, together with land use information at 1 m grid resolution. Second, we measure and analyze disparities in community social distancing by estimating variations neighborhood activity and associated patterns in community characteristics before and after the stay-at-home order. Finally, we evaluate the effect of social distancing in neighborhoods on COVID-19 infection rates and outcomes associated with localized demographic, socioeconomic, and infrastructure characteristics in order to identify disparities in health outcomes related to exposure risk. Our findings provide insight into the timely evaluation of the effectiveness of social distancing for individual neighborhoods and support a more equitable allocation of resources to support vulnerable and at-risk communities.

\section{Background}
\label{S:2}

Emerging research has begun to leverage a variety of new data resources to measure and evaluate social distancing practices. For example, the World Bank COVID-19 Mobility Task Force and Emergency Operation Center (EOC) uses anonymized Cell Detail Record (CDR) data from Mobile Network Operators (MNOs) to provide technical, operational, and decision-making support to mitigate the spread of COVID-19, particularly for data-poor countries \cite{worldbank}. Data fields include the date, number of devices, number of trips, and mean distance traveled for a given region. Specifically, the dataset contains origin--destination  information aggregated to cell tower capture areas, providing insights into movement patterns and human activity density on a given day. 

Social media, such as Twitter or Facebook, has been another popular source to measure social distancing \cite{facebook}. \citet{kayes2020automated} use Twitter data to develop an automated social distancing measurement method and sentiment detection model in Australia. The study finds that more than 80\% of Twitter users have a positive sentiment and they are willing to accept social distancing practices to protect their communities \cite{kayes2020automated}. Also, \citet{xu2020twitter} develop a social mobility index as a measure of social distancing and travel patterns by using public geotagged Twitter data. This research defines the social mobility index based on total travel distance between identified home locations and other visited locations for each user, and compares this metric before and after the pandemic declaration. The authors observe a decrease in mobility of 62\% in the U.S on average, ranging from 39\% to 77\% across different states. 

Using a range of urban sensing modalities, \citet{sensor} at the Newcastle University Urban Observatory in the United Kingdom developed a dashboard to provide social distancing status and vehicle mobility patterns in real time. This group used thousands of sensors from vehicular and pedestrian environments, such as parking lot occupancy sensors and public transportation GPS trackers, to monitor urban dynamics. They analyzed more than 1.8 billion data points to compare daily mobility patterns after the lockdown against the same day from the previous year. The research observed peaks of pedestrian movement disappeared and mobility volume decreased by more than 90\% after the stay-at-home order. Other sensing approaches include the use of computer vision and synoptic imagery to quantify pedestrian and motorized activity.

Many researchers have turned to Point-of-Interest (POI) data, which is location information primarily used by retailers for customer tracking (e.g. visitor counts) and commercial space planning, extracted from mobile phone. As POI data typically provides the location of stores or restaurants and the number of visitors, activity levels at POI locations can be used to measure social distancing at specific places. Many studies rely on POI data provided by SafeGraph, in partnership with ESRI. For example, \citet{bayhamcolorado} analyzed POI data to quantify mobility patterns in Colorado during the early stage of the COVID-19 epidemic at the state and county level. The report found that residents reduced their social activities by 80\% after the statewide stay-at-home order, and residents of higher-income counties reduced their activity levels more than lower-income counties. Other recent studies, including \citet{allcott2020polarization} and \citet{painter2020political}, use POI data to understand potential disparities in social distancing practices based on political affiliation. 

Finally, cell phone GPS data, primarily collected from smartphone applications, has gained widespread attention for its potential to understand large-scale population dynamics \cite{coven2020disparities}. Google, for instance, uses anonymized and aggregated mobility data of users based on their \textit{Google Location History} to measure mobility patterns and demonstrate the effectiveness of various social distancing policies \cite{google,wellenius2020impacts,aktay2020google}. Based on these data, research has quantified changes in the amount of time individuals spend away from their place of residence and changes in the number of visits to non-residential places, such as retail stores, parks, and transit stops, aggregated to the county level. Similarly, Apple extracts user requests for directions in Apple Maps to create anonymized, daily mobility data for metropolitan areas \cite{apple}. In addition to Google and Apple, there are third party providers that collect users' location with timestamp information from individual smartphone application and sell (or share) these data with a range of end-users researchers. As an example, the New York Times acquired data on 15 million users in the U.S. from Cuebiq, a location data provider, to quantify the number of people staying home and changes in travel patterns across the country. Similarly, Unacast has developed a social distancing score across the U.S. at the county level using their location data products, which includes an interactive dashboard \cite{unacast}.

Despite the range of data sources used to measure social distancing, many studies face nontrivial constraints caused by data limitations and methodological challenges. First, previous work is limited to relatively coarse geographical units, such as city, county, or state. In order to quantify and understand the range of effects of social distancing across socioeconomic and demographic diversity, observation of changes in activity are needed at higher spatial resolutions. Second, previous work does not account for land use classifications in determining where activity is occurring, and how those location shift over time. While POI data can capture visits to certain types of establishments, it does not, by itself, allow for measurement of activity taking place outdoors, in residential settings, or other places where check-ins may not occur. Therefore, a scalable metric of neighborhood activity change and exposure risk is necessary to understand the heterogeneous effects of social distancing on vulnerable communities, to model localized spread of disease, and to help develop more equitable and effective public health interventions targeted to at-risk communities.

\section{Data and Methods}
\label{S:3}

\begin{table}[!b]
\caption{Data Sources and descriptions}
\label{table:data}
\vspace{5mm}
\scriptsize
\def\arraystretch{1.3}
\centering
\begin{tabular}{m{2.5cm} m{1.4cm} m{2.5cm} m{5cm}}
\hline
\centering \textbf{Dataset} &  \centering \textbf{Time range} & \centering \textbf{Resolution\\(spatial/temporal)} & \centering \textbf{Source and description}  \tabularnewline \hline \hline
\centering Mobility data & \centering 2020-02-01 $\sim$ 2020-04-30 & \centering (X,Y)/second & Geotaggged data points collected from more than 200 mobile applications provided by VenPath, Inc. \\ \hline
\centering NYC Primary Land Use Tax Lot Output (PLUTO) & \centering updated 2020-02-24 & \centering Parcel/NaN & Land use and building type information provided by the NYC Department of City Planning \\\hline
\centering NYC Building Footprints & \centering updated 2020-07-06 & \centering Footprint/NaN & perimeter outline of more than 1 million buildings in NYC provided by the Department of Information Technology \& Telecommunications (DoITT). \\ \hline
\centering Road Network Data (LION) & \centering updated 2020-04-28 & \centering Street segment/NaN & Single line street base map provided by the NYC Department of Transportation \\\hline
\centering NYC COVID-19 data & \centering 2020-04-01 $\sim$ 2020-06-04 & \centering Zipcode/daily & COVID-19 confirmed cases, deaths, and positivity rates information provided by the NYC Health Department \\\hline
\centering American Community Survey (ACS) & \centering 2018 5-year estimates & \centering Zipcode/NaN & Household demographic and socioeconomic characteristics from the U.S. Census Bureau \\\hline
\centering NYC Hospital locations & \centering updated 2017-09-08 & \centering (X,Y)/NaN & List of hospitals of the NYC Health and Hospital Corporation and public hospital system \\\hline
\centering Nursing home data & \centering updated 2020-05-24 & \centering (X,Y)/NaN & Nursing home information including the number of beds and occupancy across the country provided by the Centers for Disease Control's National Healthcare Safety Network system \\ \hline
\end{tabular}
\end{table}

Our primary data are anonymized smartphone geolocations collected by VenPath, Inc. -- a data marketplace company providing mobile application data and business analytics consulting based on more than 200 various smartphone applications across the United States. The approximately 5 TB dataset covers the period from February through April 2020 and contains more than 127 billion geotagged data points associated with 120 million unique devices every month %(actual value: 184,126,018)
across the country. For this study, we extract a subset of the data falling within the Greater New York area bounding box extent ($40\degree29'46.0'' N 74\degree15'20.1'' W$ : $40\degree54'55.9'' N 73\degree42'00.0'' W$) and adjust timestamps to the Eastern Standard Time (EST) zone, resulting in 12,858,781 unique devices. After filtering for devices active for at least 14 days over the study period, the processed dataset includes 744,147 unique devices, representing approximately 8.9\% of the NYC population.

To complement our mobility data, we use a range of ancillary data as described in Table \ref{table:data} for data analysis and modeling. NYC Primary Land Use Tax Lot Output (PLUTO) data are used to obtain land use and building type information for every property in the city. The building footprint shapefile is used to identify the exact boundaries of individual buildings. NYC LION data -- a single line street base map -- are used to extract street segment geometries. We use daily NYC COVID-19 information by zipcode, which includes confirmed cases, deaths, and positive test rates, provided by the NYC Department of Health and Mental Hygiene (DOHMH). In order to contextualize neighborhood demographic, socioeconomic, housing, and public health-related characteristics, we use American Community Survey (ACS) data from the U.S. Census Bureau, NYC hospital locations from NYC OpenData, and nursing home data provided by the U.S. Centers for Disease Control (CDC). With the exception of the smartphone geolocation data, all data are publicly available and extracted from NYC or federal open data platforms. 

We explore three hypotheses. First, large-scale mobility data can represent neighborhood activity levels over time, and neighborhood social distancing can be measured by changes in this observed activity. Second, disparities in community activity changes before and after a stay-at-home order are associated with neighborhood socioeconomic and demographic characteristics. Third, variations in neighborhood social distancing result in disparities in COVID-19 infections and outcomes, controlling for differences in population health risk. 

\subsection{Building the Exposure Density Metric}
We introduce \textit{exposure density} ($E_x\rho$) as a high spatiotemporal resolution social distancing metric using large-scale mobility data without identifying and tracking individual devices. The goal of social distancing is to reduce the probability of contact between potentially infected and non-infected people; therefore, it can be defined mathematically as the inverse proportion of human activity density, represented by the number of people in a given area at a given time. Naively, a lower activity volume, holding spatial area constant, results in lower population density, thus decreasing the probability of close contacts. However, this metric needs to account for both the volume of activity in an area and the type of land use where activities occur. For example, activities in residential buildings can be a measure of people staying at home, while activities outside of residential buildings are more likely to increase exposure risk by raising the likelihood of contact with those outside of the family or household unit. Here, $E_x\rho$ is measured as the number of unique devices in a given geographical and temporal unit by land use type, specified as:
\begin{equation}
   E_x\rho = f(A_{g,t,L}) 
\end{equation}
where $g$ is the selected geographical unit (e.g. grid cell or census block group), $t$ is the temporal unit (e.g. hourly or daily), and $L$ is the land use class. 

\begin{figure}[t!]
  \scriptsize
  \centering
  \includegraphics[width=1\textwidth]{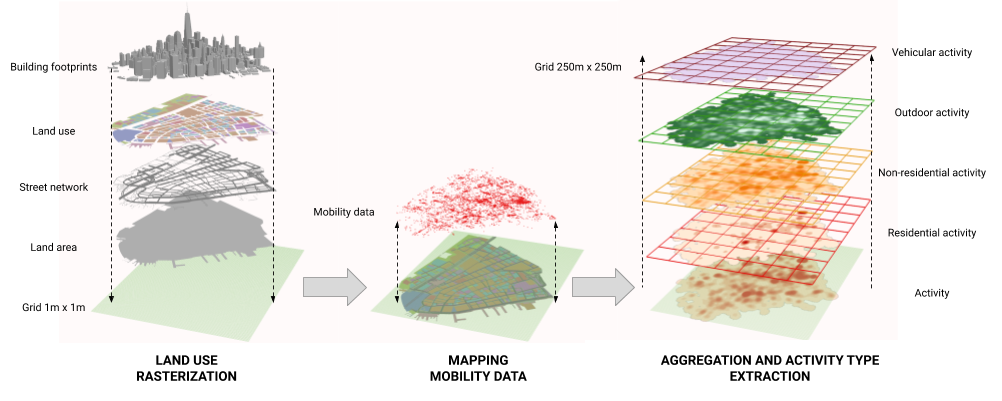}
  \caption{Visualization of data processing workflow starting from land use rasterization on 1m x 1m grid (left), spatial join of mobility data with land use information by mapping activity location on the raster (center) and aggregation of hourly activity by type on each 250m x 250m grid (right).}
  \label{figure:rasterization}
\end{figure}

In order to maintain a scalable and uniform areal unit that can be applied across different cities and regions, we divide the NYC study area into a 250 m grid (187 x 186 cells) which we use for aggregation of the mobility data. To enrich the mobility data with land use information, we create a 1 m resolution raster with the extents and the coordinate system matching the aforementioned 250 m grid. The land use raster combines the geographical city limits and land use classification derived from PLUTO data, together with street and sidewalk boundaries and building footprints for more than 1,000,000 buildings. Each category of land cover is then classified by an integer (e.g. 10 for residential property, 50 for outdoor open space, and so on). Each 1 m cell is therefore identified by its index, location and associated land cover category. This allows us to assign each geolocation data point from the mobility dataset to a specific land use cell. 
To estimate population density, we count the hourly number of unique devices by each 250 m grid cell and the corresponding land use category based on the raster cell. Our data processing workflow is visualized in Figure \ref{figure:rasterization}. The rasterization process was implemented in Python and deployed on NYU Center for Urban Science and Progress' Research Computing Facility and the activity computation was performed with PySpark on a Hadoop distributed computing cluster using NYU's High Performance Computing platform.

Our 250 m grid cell level measurement can be aggregated into larger geospatial units in order to estimate neighborhood activities at different scales. In this work, we use zipcode aggregation to compute neighborhood activity to align with COVID-19 infection data provided by the DOHMH. The zipcode aggregated ($E_x\rho$) is defined as:
\begin{equation}
    A_{z,g,t,L} = \frac{1}{N_{z}}\sum_{i=1}^{N_{z}} A_{g,t,L}
\end{equation}
where $A_{z,g,t,L}$ is the average number of hourly unique devices in a 250m$\times$250m grid cell by land use type $L$ in a given zipcode $z$, and $N_{z}$ is the number of grid cells in zipcode $z$. Based on our social distancing metric, changes in mobility activity by residential, non-residential, and outdoor land uses in a neighborhood over the study time period are examined. We filter out activities from major roads used exclusively by motor vehicles (those without sidewalks or pedestrian access) to remove vehicular activity within a given neighborhood.

\subsection{Analyzing disparities in exposure density}
To understand disparities in exposure density and behavioral responses to social distancing mandates across neighborhoods, we apply an unsupervised machine learning clustering algorithm based on a \textit{pre/post} comparative analysis. We extract ($E_x\rho$) subsets for two, two-week periods, defined as the \textit{pre-impact} period (February 16 through February 29 2020) and the \textit{impact} period (March 29 through April 11 2020), to measure changes in ($E_x\rho$) before and after the state-mandated stay-at-home order. In order to take into account both the absolute change in activity volume and the change in the proportion of activity type, we create six (6) input variables for the zipcode clustering analysis specified as:
\begin{equation}
    \begin{array}{c c c}
      A_{change,z,L} = \frac{A_{post,z,L}-{A_{pre,z,L}}}{A_{pre,z,L}} \\ \\
    P_{change,z,L} = \frac{P_{post,z,L}-{P_{pre,z,L}}}{P_{pre,z,L}} \\ \\
    (L  \in  \mbox{residential, non-residential, and outdoor})
    \end{array}
\end{equation}
where $A_{change,z,L}$ is average hourly activity volume change for residential, non-residential, and outdoor land uses in zipcode $z$ based on the pre-impact period activity level ($A_{pre,z,L}$) and the impact period level ($A_{post,z,L}$). $P_{change,z,L}$ is the average hourly change in activity based on the proportion of those activities occurring in different land use types. Neighborhood activity by land use classification is defined as the proportion of activity in a given land use (residential, non-residential, and outdoors) grid cell.

To identify similarities in the change in ($E_x\rho$) , we use a Wards' metric-based agglomerative clustering algorithm, which a widely-used bottom-up hierarchical unsupervised clustering methods. It begins with each data point considered as an individual cluster. At each iteration, the closest two clusters merge with each other based on the proximity matrix measured by Euclidean distance until all data points form a single cluster \cite{hastie2009elements}. Input data is in the form of a 177$\times$6 vector -- 177 zipcode neighborhoods and 6 features -- and the optimized number of clusters is determined by the corresponding dendrogram (hierarchical tree diagram) based on the similarities and dissimilarities of the objects. This clustering process is specified as:
\begin{equation}
    \Delta(C_{i},C_{j}) = \sum_{k \in C_{i} \cup C_{j}}^{} \|\vec{x}_{k} - \vec{m}_{C_{i} \cup C_{j}}\|^{2} - \sum_{k \in C_{i}}^{} \|\vec{x}_{k} - \vec{m}_{C_{i}}\|^{2} - \sum_{k \in C_{j}}^{} \|\vec{x}_{k} - \vec{m}_{C_{j}}\|^{2}
\end{equation}
where $\Delta(C_{i},C_{j})$ is a merging cost of combining clusters $C_{i}$ and $C_{j}$ (distance between clusters), $\vec{m}_{C}$ is the centroid of cluster $C$, and $\vec{x}_{k}$ is an individual element within a cluster. The resultant clustered neighborhood groups are then integrated with demographic and socioeconomic characteristics, housing and urban form features, and COVID-19 infection and outcome data. By using a one-way ANOVA (analysis of variance) test and a Tukey's test for post-hoc analysis, we identify statistically significant differences between classified groups regarding behavioral responses and associated neighborhood characteristics.

\subsection{Identifying the impact of exposure density and neighborhood behavior change on infectivity}
In order to evaluate the effect of neighborhood behavior changes on COVID-19 infection rates for the 177 neighborhoods included in the study, we first measure Pearson correlation coefficients for observed community activity changes before and after the stay-at-home order and disease infection case rates -- daily new confirmed cases per 100,000 people and cumulative cases per 100,000 people -- while accounting for an incubation period. 

\begin{table}[!b]
\caption{Description of COVID-19 infection rates}
\label{table:infection_rates}
\vspace{5mm}
\footnotesize
\def\arraystretch{1.1}
\centering
\begin{tabular}{m{3cm} m{9.5cm}}
\hline
\centering \textbf{Infection rate indicators} &  \centering \textbf{Description}  \tabularnewline \hline \hline
\centering Case rate & Number of confirmed cases per 100,000 by zipcode  \\ 
\centering Death rate & Number of confirmed deaths per 100,000 by zipcode \\ 
\centering Positivity rate & Percentage of positive tests using a polymerase chain reaction (PCR) test  \\ 
\centering Deaths per case & Number of deaths per confirmed case  \\ \hline
\end{tabular}
\end{table}

\begin{table*}[t!]
\caption{Summary statistics of input variables}
\label{table:input_variables}
\vspace{3mm}
\tiny
\def\arraystretch{1.1}
\centering
\begin{tabular}{m{4cm} m{6cm} m{2.5cm}}
\hline
\textbf{Variable} & \centering \textbf{Description} & \centering \textbf{Statistics}  \tabularnewline \hline \hline
Exposure density change & \centering \% change in activities outside of residential buildings & \centering -0.18 (0.20)  \tabularnewline
Neighborhood clusters & \centering binary variables for each identified cluster & \centering - \tabularnewline
White & \centering \% non-Hispanic White population & \centering 0.47 (0.27)  \tabularnewline
Black & \centering \% Black population & \centering 0.21 (0.24)  \tabularnewline
Hispanic & \centering \% Hispanic population & \centering 0.26 (0.20)  \tabularnewline
Asian & \centering \% Asian population & \centering 0.15 (0.14)  \tabularnewline
Age group 25-34 & \centering \% of population 25-34 years old  & \centering 0.18 (0.06) \tabularnewline
Age group over 65 & \centering  \% of population over 65 years old & \centering 0.14 (0.05) \tabularnewline
Household size & \centering Average household size in zipcode & \centering 2.64 (0.50)  \tabularnewline
Household with children & \centering \% of households with children under 18 & \centering  0.25 (0.08) \tabularnewline
Educational attainment & \centering \% of population with Bachelor's degree & \centering  0.23 (0.10) \tabularnewline
No health insurance & \centering \% of households without health insurance & \centering 0.08 (0.04)  \tabularnewline
Public health insurance & \centering \% of households with public insurance & \centering 0.39 (0.14)  \tabularnewline
Commute time & \centering Average commute time (minutes) & \centering 40.76 (7.12)  \tabularnewline
Median income & \centering Median income in zipcode & \centering  74K (37K) \tabularnewline
Unemployment rate & \centering \% of labor force unemployed & \centering 0.07 (0.03) \tabularnewline
Owner occupied units & \centering \% of housing units occupied by owner & \centering  0.37 (0.22) \tabularnewline
One or two family home & \centering \% of one or two family home units & \centering  0.30 (0.31)  \tabularnewline
Public housing & \centering \% of public housing units & \centering 0.05 (0.08)  \tabularnewline
Residential area & \centering Residential building area as \% of total built area & \centering  0.65 (0.20) \tabularnewline
Office area & \centering Office building area as \% of total built area & \centering    0.09 (0.15)\tabularnewline
Commercial area & \centering Commercial (non-office) building area as \% of total built area & \centering 0.28 (0.17) \tabularnewline
Hospital & \centering Hospital(s) located in zipcode (yes=1) & \centering 0.21 (0.41) \tabularnewline
Nursing home & \centering Number of occupied nursing home beds & \centering 507 (837) \tabularnewline

\hline
\multicolumn{3}{m{10cm}}{\textit{Note: } Standard deviations are in parentheses for continuous variables with mean values.}

\end{tabular}
\end{table*}

Then, we develop bivariate and multivariate log-transformed regression models to identify any statistical significant effect of $E_x\rho$ on infections, controlling for neighborhood characteristics. Ordinary least squares (OLS) regression models are applied with four (4) dependent variables, representing four measures of COVID-19 infectivity (Table \ref{table:infection_rates}), including case rate, death rate, positivity rate, and deaths per case.  Table \ref{table:input_variables} provides descriptive statistics for the included independent variables. The bivariate models take $E_x\rho$ change (as a percent) as a continuous variable to measure the marginal effects of activity change on infection rates. The multivariate models, on the other hand, use dummy variables for each clustered neighborhood group to evaluate disparities between groups. The linear models are specified as:
\begin{equation}
    \begin{array}{c c}
    y = \beta_{0} + \beta_{1} X_{1} + \varepsilon  \mbox{          (Bivariate model)}  \\ \\
    y = \beta_{0} + \sum_{i=1}^{n} \beta_{i} X_{i} + \varepsilon  \mbox{          (if $n>1$, multivariate model)}
    \end{array}
\end{equation}
where $y$ is the logarithmic transformed zipcode-level COVID-19 outcome variable, $X_{1}$ for the bivariate model is exposure density change, $X_{i} (i>1)$ for the multivariate model includes the cluster group dummy variables the set of neighborhood demographic, socioeconomic, and built environment features, and $\varepsilon$ is the error term. We also consider interaction terms between the neighborhood groups and other social determinants of health. We use correlation tests and Variance Inflation Factors (VIFs) analysis to identify multicollinearity as part of the feature selection process. The coefficients $\beta_{i}$ evaluate the effects of neighborhood $E_x\rho$ on disparities in disease risk.

\section{Results and Findings}
\label{S:4}

\subsection{Changes in exposure density by neighborhood}

\begin{figure}[b!]
  \scriptsize
  \centering
  \includegraphics[width=1\textwidth]{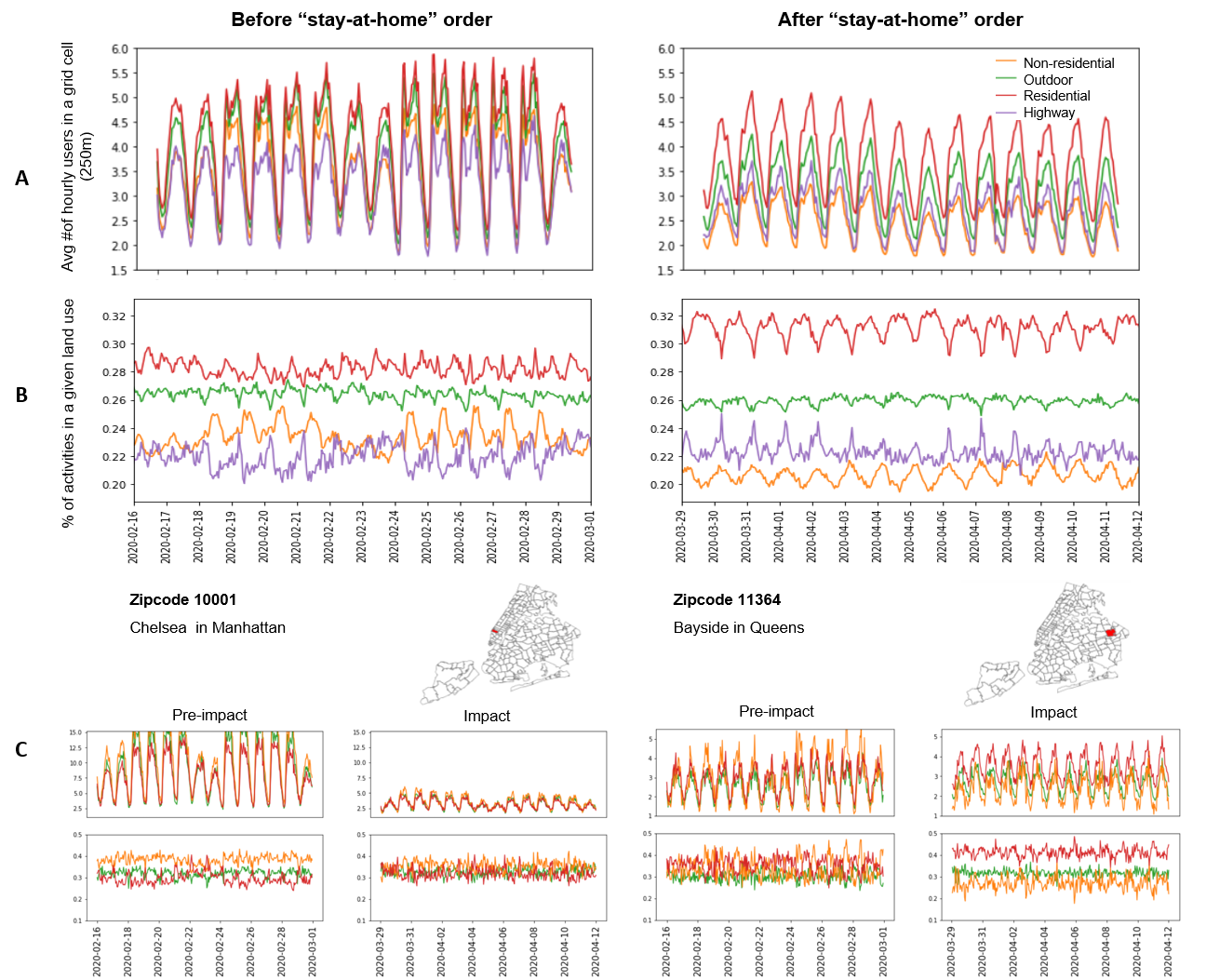}
  \caption{Neighborhood activity measurements by land use types at 250m grid cell level before and after the stay-at-home order (A) citywide average neighborhood activity volume changes (B) citywide average neighborhood activity proportion changes (C) Example neighborhoods - zipcode 10001 (Chelsea, Manhattan) and zipcode 11364 (Bayside, Queens)}
  \label{figure:overall_activity_change}
\end{figure}

\begin{figure}[b!]
  \scriptsize
  \centering
  \includegraphics[width=1\textwidth]{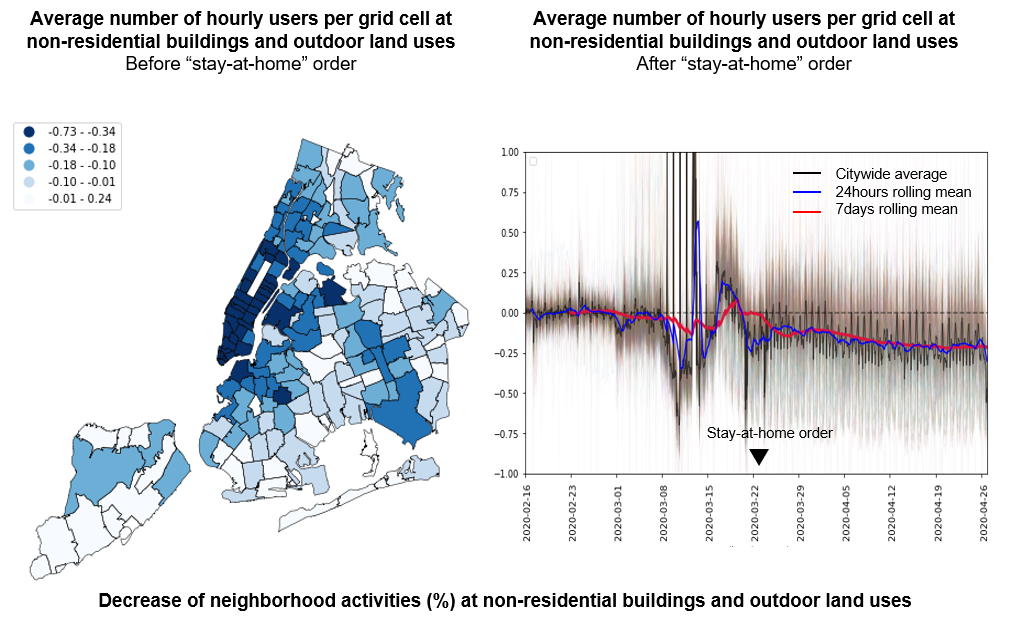}
  \caption{Neighborhood exposure density. Left: Percentage change in exposure density by zipcode. Right: Individual time series representing exposure density change over time by zipcode with citywide moving average reference. Citywide, there is an approximately 20\% decrease in exposure density after the stay-at-home order.}
  \label{figure:ngbg_activities}
\end{figure}

Citywide and neighborhood hourly activity change by land use type are presented in Figure \ref{figure:overall_activity_change}. The citywide overall activity volume (Figure \ref{figure:overall_activity_change}A) decreased, especially outdoors (sidewalks or parks) and in non-residential spaces (office and retail), which show significant reductions (20\% and 33\% decrease, respectively) after the stay-at-home order when compared to the pre-COVID baseline. We also observe changes to typical activity peak periods before and after the stay-at-home order. In the pre-COVID period, there are two peaks (around 8am and 6pm) each weekday, a result of a commute patterns during rush hour periods. The morning peak, however, is no longer discernible in the post-COVID period as remote work and school closures reduce commuting activity. In addition to population volume changes, the analysis of where activity is occurring reveals significant behavior changes across the City (Figure \ref{figure:overall_activity_change}B). We observe residential activities increased by three percentage points (a +10\% change) and non-residential activities decreased by three percentage points (a -13\% change) after the stay-at-home order. This demonstrates both a decrease in overall population, as many residents left the city, and a shift in activities from non-residential and outdoor areas to residential buildings for those that remained. Figure \ref{figure:overall_activity_change}C illustrates this change in activity (both volume and proportion) for two exemplar neighborhoods. In the case of zipcode 10001 in Midtown Manhattan, the number of people staying in the neighborhood dramatically decreased by more than 60\% with activities of those who remain becoming more evenly distributed between residential, non-residential, and outdoor land uses. This highlights the exodus of residents from the city and the reduction in the number of visitors to the neighborhood. Of those who stayed, more remained at home than would be expected. Another example neighborhood -- zipcode 11364 at the eastern edge of Queens -- shows a substantial increase in residential activity volume and in the proportion of activities occurring in residential areas. 

As transmission risk increases with a greater probability of close contacts outside of the household or family unit, we quantify $E_x\rho$ based on activities in non-residential buildings and outdoor areas. We measure the average number of hourly users per grid cell outside of residential buildings for 177 zipcodes during the pre-impact period and after stay-at-home order. Figure \ref{figure:ngbg_activities} illustrates the percentage change in $E_x\rho$. There is a 20\% citywide decrease in $E_x\rho$ after the stay-at-home order, but we observe significant disparities in behavior change across neighborhoods (see Figure \ref{figure:ngbg_activities}). For instance, there is a 50\% decrease, on average, in $E_x\rho$ in Manhattan, Long Island City, and Downtown Brooklyn, while activities in the south area of Staten Island, South Brooklyn, and the east side of Queens exhibit only minor variations. 

\subsection{Disparities in neighborhood exposure density}
\begin{figure*}[t!]
  \scriptsize
  \centering
    \includegraphics[width=0.9\textwidth]{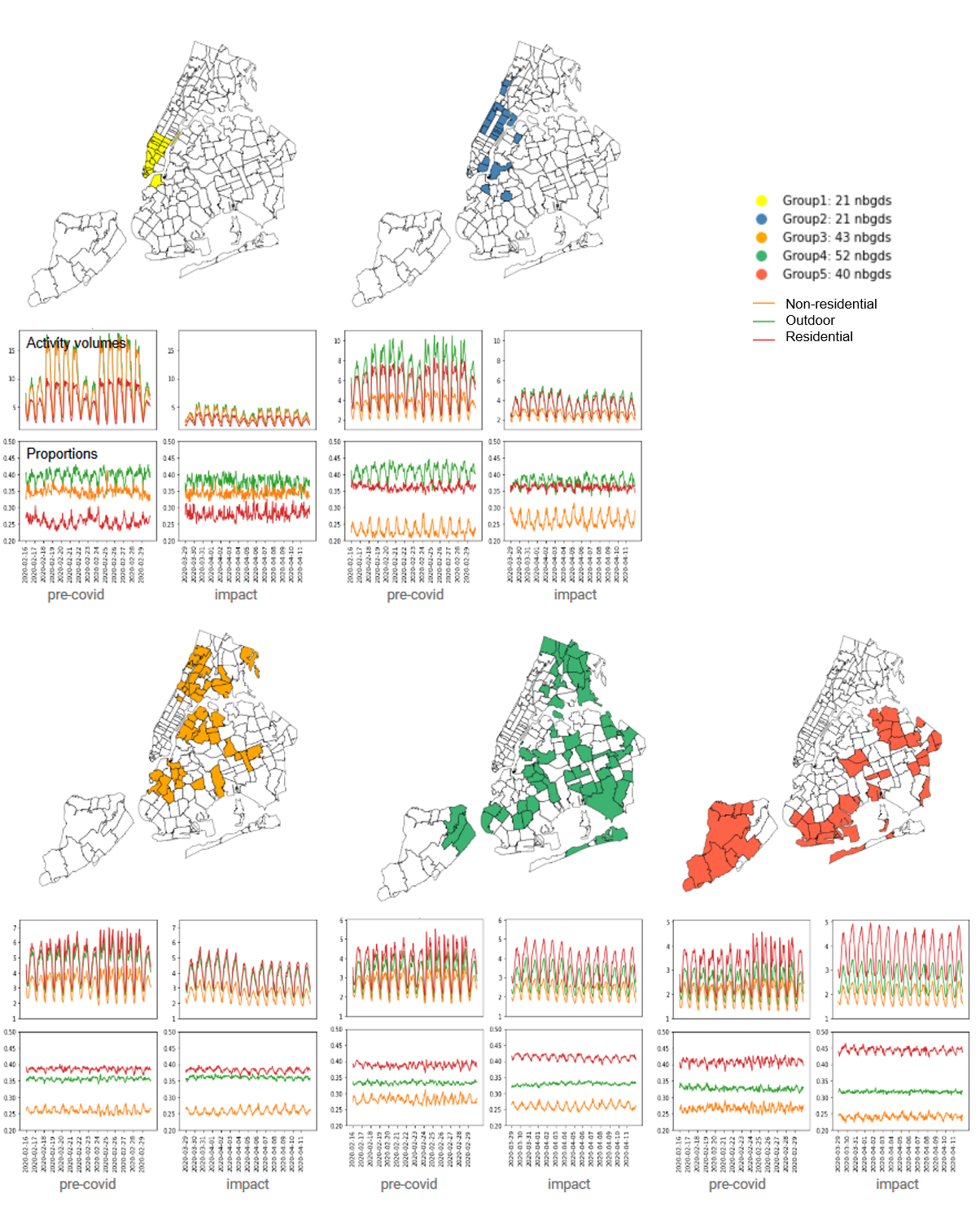}
    \caption{Spatial patterns of the agglomerative clustering results and associated neighborhood activity change (top time series: activity volumes by land use, bottom time series: activity proportions by land use)}
    \label{fig:clustering_result_separate}
\end{figure*}

We classify neighborhoods into distinct groups using a hierarchical clustering algorithm based on changes in community activity levels and proportions before and after the stay-at-home order. Figure \ref{fig:clustering_result_separate} illustrates the spatial patterns of the clustering output with associated time series of neighborhood activity and where (by land use type) that activity is occurring. Descriptive statistics of input variables and neighborhood features for each group, shown in Table \ref{table:clustering_inputs} and Table \ref{table:clustering_features} respectively, reveal distinct neighborhood profiles based on changes in $E_x\rho$ over time.

Based on this analysis, we identify five (5) neighborhood clusters. Group 1 (21 zipcodes) and Group 2 (21 zipcodes), which we label ''outflow" neighborhoods, are primarily located in Manhattan and Downtown Brooklyn and represent substantial changes in $E_x\rho$ after the stay-at-home order. As shown in Table \ref{table:clustering_inputs}, the average activity volume change for Group 1 and Group 2 is -56.5\% and -33.5\%, respectively, meaning these two neighborhood groups experienced nontrivial declines in normal activity levels -- across all land use types -- during the pandemic. Most neighborhoods in Group 1 and Group 2 have a higher percentage of younger, non-Hispanic white residents, relatively smaller average household size, and higher incomes and educational attainment. This indicates residents in these clusters are among the least vulnerable population groups and can afford to leave their home neighborhoods (or stay at home) by shifting to remote working environments to avoid the exposure risk, resulting in a decrease exposure density. Even though these two clusters present similar outflow patterns with respect to neighborhood activity volume, the activity proportion changes exhibit some notable differences. While the proportion of residential activities in Group 1 increased by 12\% without any significant changes in non-residential and outdoor activities, Group 2 shows a 14\% increase in non-residential activity, a function of the relative pre-COVID resident population size. Therefore, we refine the labels for Group 1 and Group 2 as ``outflow-mixed use" and ``outflow-residential", respectively. 
\begin{table*}[t!]
\caption{Descriptive statistics of neighborhood clusters}
\label{table:clustering_inputs}
\vspace{3mm}
\tiny
\def\arraystretch{1.0}
\centering
\begin{tabular}{m{4.3cm} m{1.4cm} m{1.4cm} m{1.4cm} m{1.4cm} m{1.4cm}}
\hline
% \centering \textbf{Neighborhood\\activity change (\%)} & \centering \textbf{Group 1\\``Stable-stable" \\ (green)} & \centering \textbf{Group 2 \\``Outflow-mixed"\\(yellow)} & \centering \textbf{Group 3\\``Influx"\newline  \\ (red)} & \centering \textbf{Group 4 \\``Outflow-residential" \\ (blue)} & \centering \textbf{Group 5\\``Outflow-stable" \\ (orange)} \tabularnewline \hline \hline
\centering \textbf{Neighborhood\\activity change (\%)} & \centering \textbf{Group 1 \\``Outflow-mixed"\\(yellow)} & \centering \textbf{Group 2 \\``Outflow-residential" \\ (blue)} & \centering \textbf{Group 3\\``Outflow-stable" \\ (orange)}& \centering \textbf{Group 4\\``Stable-stable" \\ (green)}  & \centering \textbf{Group 5\\``Shelter-in-place"\newline  \\ (red)}   \tabularnewline \hline \hline
\centering Residential volume & \centering -0.52 & \centering -0.37 & \centering -0.20 & \centering -0.01 & \centering 0.20  \tabularnewline
\centering Residential proportion & \centering 0.12 & \centering 0.01 & \centering -0.01 & \centering 0.07 & \centering 0.09  \tabularnewline
\centering Non-residential volume & \centering -0.60 & \centering -0.28 & \centering -0.19 & \centering -0.13 & \centering -0.00  \tabularnewline
\centering Non-residential proportion & \centering -0.01 & \centering -0.14 & \centering 0.00 & \centering -0.07 & \centering -0.09  \tabularnewline
\centering Outdoor volume & \centering -0.61 & \centering -0.42 & \centering -0.18 & \centering -0.07 & \centering 0.07  \tabularnewline
\centering Outdoor proportion & \centering -0.04 & \centering -0.07 & \centering 0.02 & \centering -0.01 & \centering -0.03  \tabularnewline
\hline
\end{tabular}
% \end{table*}
\vspace{5mm}
% \begin{table*}[b!]
\caption{Neighborhood cluster characteristics.
Statistically significant differences between groups based on one-way ANOVA and Tukey’s multi-comparison method. Mean values with standard deviation in parentheses}
\label{table:clustering_features}
\vspace{3mm}
\tiny
\def\arraystretch{1.0}
\centering
\begin{tabular}{m{3.8cm} m{1.5cm} m{1.5cm} m{1.5cm} m{1.5cm} m{1.5cm}}
\hline
% \centering \textbf{Feature} & \centering \textbf{Group 1\\``Stable-stable" \\ (green)} & \centering \textbf{Group 2 \\``Outflow-mixed"\\(yellow)} & \centering \textbf{Group 3\\``Influx"\newline  \\ (red)} & \centering \textbf{Group 4 \\``Outflow-residential" \\ (blue)} & \centering \textbf{Group 5\\``Outflow-stable" \\ (orange)} \tabularnewline \hline \hline

\centering \textbf{Feature} & \centering \textbf{Group 1 \\``Outflow-mixed"\\(yellow)} & \centering \textbf{Group 2 \\``Outflow-residential" \\ (blue)} & \centering \textbf{Group 3\\``Outflow-stable" \\ (orange)}& \centering \textbf{Group 4\\``Stable-stable" \\ (green)}  & \centering \textbf{Group 5\\``Shelter-in-place"\newline  \\ (red)}   \tabularnewline \hline \hline
Demographic and \\ socioeconomic features &  &  &  &  &  \tabularnewline
\hspace{2mm}Age group 25-34 (\%)    & \raggedleft 0.28 (0.08) & \raggedleft 0.22 (0.05) & \raggedleft 0.19 (0.04) & \raggedleft 0.16 (0.06) & \raggedleft 0.13 (0.02)  \tabularnewline
\hspace{2mm}Age group over 65 (\%)  & \raggedleft 0.12 (0.07) & \raggedleft 0.15 (0.06) & \raggedleft 0.12 (0.04) & \raggedleft 0.14 (0.05) & \raggedleft 0.17 (0.07)  \tabularnewline
\hspace{2mm}Black (\%)              & \raggedleft 0.05 (0.04) & \raggedleft 0.14 (0.17) & \raggedleft 0.27 (0.22) & \raggedleft 0.31 (0.28) & \raggedleft 0.16 (0.25)  \tabularnewline
\hspace{2mm}Non Hispanic (\%)       & \raggedleft 0.90 (0.05) & \raggedleft 0.77 (0.19) & \raggedleft 0.60 (0.21) & \raggedleft 0.72 (0.07) & \raggedleft 0.82 (0.11)  \tabularnewline
\hspace{2mm}Foreign born (\%)       & \raggedleft 0.16 (0.08) & \raggedleft 0.14 (0.05) & \raggedleft 0.18 (0.08) & \raggedleft 0.15 (0.07) & \raggedleft 0.13 (0.08)  \tabularnewline
\hspace{2mm}Avg household size      & \raggedleft 1.92 (0.26) & \raggedleft 2.21 (0.35) & \raggedleft 2.61 (0.31) & \raggedleft 2.90 (0.45) & \raggedleft 2.91 (0.37)  \tabularnewline
\hspace{2mm}College degree (\%)     & \raggedleft 0.40 (0.07) & \raggedleft 0.31 (0.09) & \raggedleft 0.20 (0.08) & \raggedleft 0.19 (0.07) & \raggedleft 0.20 (0.04)  \tabularnewline
\hspace{2mm}Unemployment rate   & \raggedleft 0.04 (0.01) & \raggedleft 0.05 (0.03) & \raggedleft 0.08 (0.03) & \raggedleft 0.08 (0.04) & \raggedleft 0.06 (0.02)  \tabularnewline
\hspace{2mm}Healthcare support workers (\%) & \raggedleft 0.01 (0.01) & \raggedleft 0.03 (0.02) & \raggedleft 0.06 (0.04)  & \raggedleft 0.07 (0.04) & \raggedleft 0.05 (0.03)  \tabularnewline
\hspace{2mm}Retail service workers (\%) & \raggedleft 0.03 (0.01) & \raggedleft 0.04 (0.02) & \raggedleft 0.06 (0.01) & \raggedleft 0.05 (0.02) & \raggedleft 0.05 (0.02)  \tabularnewline
\hspace{2mm}Median Income (\$)      & \raggedleft 133K & \raggedleft 90K & \raggedleft 54K & \raggedleft 62K & \raggedleft 72K  \tabularnewline
\hspace{2mm}Avg commute time (minute) & \raggedleft 27.05 (3.00) & \raggedleft 33.83 (4.15) & \raggedleft 41.86 (3.23) & \raggedleft 44.7 (3.87) & \raggedleft 45.30 (3.73)  \tabularnewline
\hspace{2mm}No health insurance (\%) & \raggedleft 0.04 (0.02) & \raggedleft 0.06 (0.03) & \raggedleft 0.09 (0.03) & \raggedleft 0.09 (0.04) & \raggedleft 0.07 (0.04)  \tabularnewline
\hspace{2mm}Owner occupied units (\%) & \raggedleft 0.26 (0.12) & \raggedleft 0.23 (0.12) & \raggedleft 0.22 (0.14) & \raggedleft 0.41 (0.21) & \raggedleft 0.59 (0.20)  \tabularnewline
Urban form features &  &  &  &  &  \tabularnewline
\hspace{2mm}Residential area (\%)   & \raggedleft 0.30 (0.20) & \raggedleft 0.71 (0.13) & \raggedleft 0.69 (0.14) & \raggedleft 0.69 (0.14) & \raggedleft 0.71 (0.18)  \tabularnewline
\hspace{2mm}Office area (\%)        & \raggedleft 0.43 (0.24) & \raggedleft 0.05 (0.06) & \raggedleft 0.05 (0.03) & \raggedleft 0.04 (0.03) & \raggedleft 0.03 (0.02)  \tabularnewline
\hspace{2mm}Commercial area (\%)    & \raggedleft 0.57 (0.22) & \raggedleft 0.24 (0.10) & \raggedleft 0.25 (0.12) & \raggedleft 0.25 (0.13) & \raggedleft 0.21 (0.13)  \tabularnewline
\hspace{2mm}One or two family units (\%) & \raggedleft 0.00 (0.00) & \raggedleft 0.03 (0.05) & \raggedleft 0.15 (0.15) & \raggedleft 0.41 (0.27) & \raggedleft 0.64 (0.26)  \tabularnewline
COVID-19 features &  &  &  &  &  \tabularnewline
\hspace{2mm}Case rate       & \raggedleft 1166.60 (431.88) & \raggedleft 1570.96 (621.38) & \raggedleft 2475.90 (786.84) & \raggedleft 2790.36 (777.17) & \raggedleft 2534.96 (630.57)  \tabularnewline
\hspace{2mm}Death rate      & \raggedleft 91.12 (76.79) & \raggedleft 150.63 (84.10) & \raggedleft 219.87 (83.11) & \raggedleft 224.46 (97.73) & \raggedleft 195.78 (116.87)  \tabularnewline
\hspace{2mm}Positivity rate & \raggedleft 0.11 (0.03) & \raggedleft 0.15 (0.05) & \raggedleft 0.22 (0.05) & \raggedleft 0.24 (0.04) & \raggedleft 0.23 (0.04)  \tabularnewline
\hline
\end{tabular}
\end{table*}

Group 3 (43 zipcodes) neighborhoods exhibit a 19\% decrease, on average, in exposure density. This is driven by a reduction in population density from those leaving the city. These neighborhoods maintain a stable proportion between the different land uses, indicating that the residents who remain in these communities maintain their regular behavior patterns. When compared to the ``outflow" groups, these ``stable-outflow" communities have higher proportions of racial and ethnic minorities, foreign born residents, lower median income, as well as significantly higher proportions of renter households and those without health insurance. Additionally, a greater percentage of employees in these neighborhoods work in retail services and healthcare support occupations, essential businesses that were not required to close during the outbreak. Like Group 3 neighborhoods, communities in the Group 4 cluster have stable activity patterns over time; however, these neighborhoods did not see a significant out-mover population. These communities, which we label ``stable-stable", are comprised of socioeconomically vulnerable households and the highest proportion of racial minorities, coupled with the second lowest income, large average household size, high unemployment rate, low educational attainment, and a large share of healthcare support workers. Such socially and economically vulnerable neighborhoods are less likely to be able to work from home as the character of their occupation requires physical presence at the workplace, leading to fewer opportunities to reduce exposure to others. We also find that the relatively modest change in exposure density in these ``stable" groups (18\% and 10\% decrease in non-residential activity density for Group 3 and Group 4, respectively) is associated with significantly higher infection rates. Particularly, the ``stable-stable" neighborhood group shows the highest case rate (2790.36), death rate (224.46), and positivity rate (0.24) in the City. 

In comparison to other clusters, Group 5 (``shelter-in-place") neighborhoods demonstrate a 20\% increase in local activity volume for residential activities and a 7\% increase for outdoor activities. In addition to increasing neighborhood activity volume, residents staying in these neighborhoods are found to shift activity to residential buildings (by 10\%) and away from non-residential and outdoor activities (by 6\%). While non-residential activities are found to decrease as a proportion of the three activity types, the increase in the overall volume of activity leads to a net increase in exposure density. This group has the highest proportion of elderly population, the largest household size, moderate incomes, a relatively lower percentage of racial and ethnic minorities, and a significantly higher homeownership rate. This indicates that activity in these neighborhoods, where population density is the lowest in the city, became more localized. As a result, Group 5 experienced the second-highest infection rate (2534.96 case rate) in the city. 

\subsection{Disparities in health outcomes: Effects of exposure density on infection rates}

\begin{figure}[t!]
  \scriptsize
    \centering \includegraphics[width=1\textwidth]{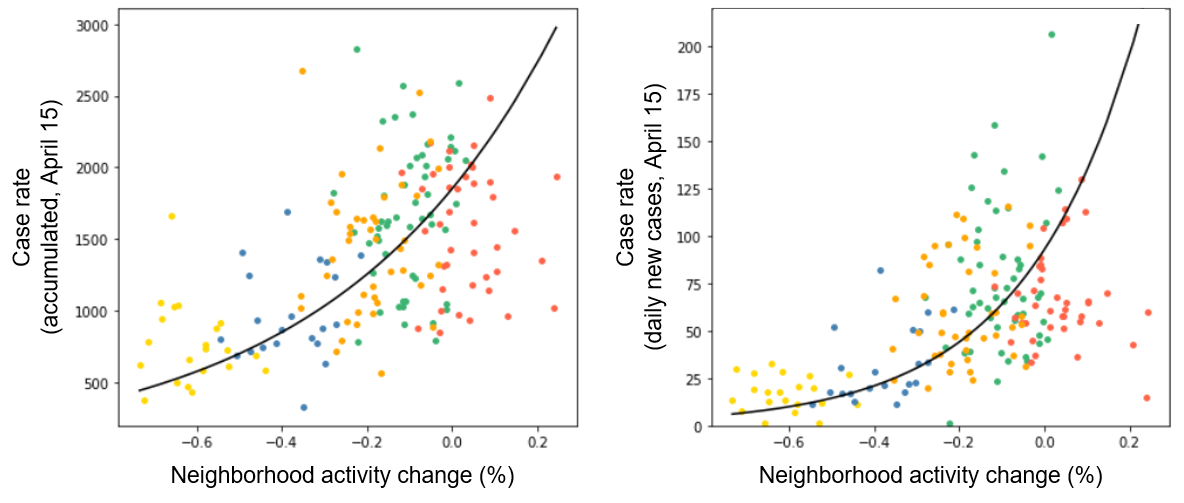}
    \caption{Scatter plots of exposure density versus COVID-19 case rates,  with incubation period. Left: cumulative cases per 100,000 people, Right: daily new confirmed cases on April 15. Colors represent individual clusters and black curves are exponential best-fit lines.}
    \label{fig:correlation}
\end{figure}

\begin{figure}[b!]
  \centering
    \includegraphics[width=1\textwidth]{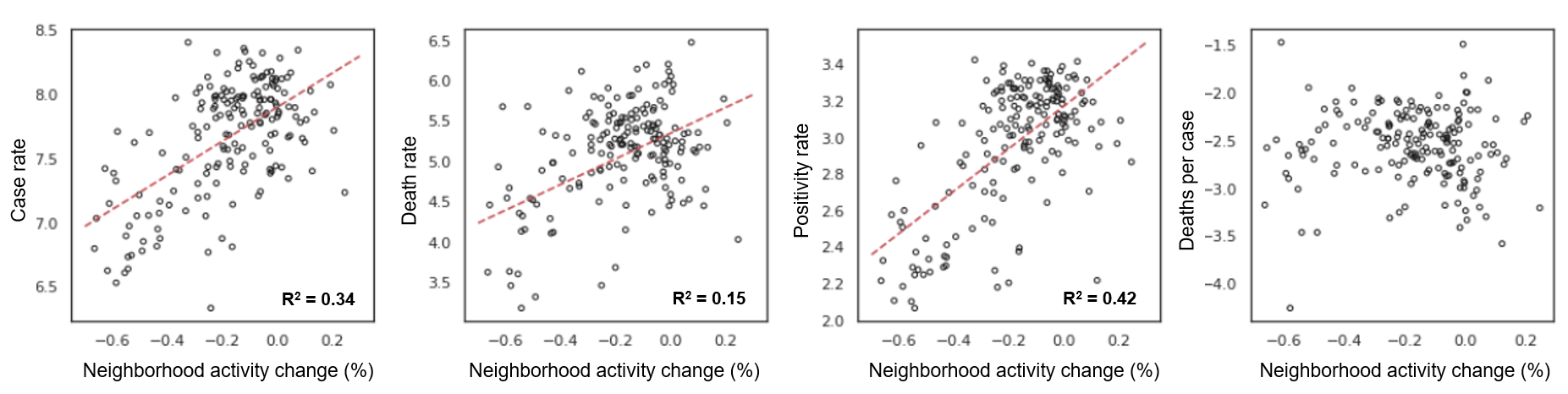}
    \caption{Scatter plot of exposure density versus the log--transformed cumulative COVID-19 1) case rate, 2) death rate, 3) positivity rate, and 4) deaths per case -- with linear best fit lines.}
    \label{fig:bivariate_modeling}
\end{figure}

The number of confirmed cases per 100,000 people (daily new cases and accumulated cases) on April 15 (five (5) days after the impact period to account for the virus incubation period) is plotted against the net change in exposure density before and after the social distancing order, as shown in Figure \ref{fig:correlation}. We observe an exponential relationship based on the scatter plots with statistically significant positive correlations between exposure density and infection rates ($\rho$ = 0.52 and $\rho$ = 0.47, respectively). 

The results of the bivariate regression model are shown in Figure \ref{fig:bivariate_modeling}. Exposure density is found to be correlated with most infection metrics, namely case rate, death rate, and positivity rate, while, as expected, not being a statistically significant determinant of mortality rate. These three models explain 34\%, 15\%, and 42\% of the variance in outcomes, respectively, with the highest explanatory power exhibited in the positivity rate model. According to the model outputs, a one percentage point decrease in exposure density is associate with a 1.33\% reduction in case rate, a 1.59\% in death rate, and a 1.16\% decrease in positivity rate in NYC. Based on these results, if all neighborhoods reduced activities away from home by 10\% as compared to normal activity levels prior to the stay-at-home order, the City could have avoided more than 28,000 COVID-19 cases and saved 2,000 lives.

\begin{table*}[t!]
\caption{Results of the multivariate regression models} 
\label{table:multivarite_model}
\vspace{3mm}
\tiny
\def\arraystretch{1.3}
\centering
\begin{tabular}{m{3.5cm} m{2cm} m{2cm} m{2cm} m{2cm}}
\hline
 & \centering \textbf{Model 1:\\case rate} & \centering \textbf{Model 2:\\death rate} & \centering \textbf{Model 3:\\positivity rate}  & \centering \textbf{Model 4:\\deaths per case}  \tabularnewline \hline \hline
 & \centering Num of obs.=177 & \centering Num of obs.=177 & \centering Num of obs.=177 & \centering Num of obs.=177 \tabularnewline
  & \centering F-stats.=35.69 & \centering F-stats.=16.59 & \centering F-stats.=53.26 & \centering F-stats.=10.90 \tabularnewline
    & \centering Prob$>$F=0 & \centering Prob$>$F=0 & \centering Prob$>$F=0 & \centering Prob$>$F=0 \tabularnewline
Feature    & \centering R-squared=0.77 & \centering R-squared=0.61 & \centering R-squared=0.83 & \centering R-squared=0.50 \tabularnewline \hline
Intercept & 7.040(0.171)*** & 2.862(0.403)*** & 2.359(0.116)*** & -3.848(0.249)*** \tabularnewline
Group ``outflow-mixed" and ``outflow-residential" & -.0632(0.135)*** & -0.716(0.318)*** & -0.443(0.091)*** & 0.128(0.197) \tabularnewline
Group ``outflow-stable" & -0.436(0.142)*** & -0.003(0.335) & -0.228(0.096)** & 0.426(0.207)** \tabularnewline
Group ``influx" & -0.051(0.115) & -0.010(0.273) & -0.130(0.078)* & 0.050(0.169) \tabularnewline
\% Black & 0.005(0.001)*** & 0.007(0.002)*** & 0.004(0.001)*** & 0.002(0.001)* \tabularnewline
\% Hispanic & 0.009(0.001)*** & 0.003(0.003)*** & 0.005(0.001)*** & 0.006(0.002)* \tabularnewline
\% units occupied by owner & 0.002(0.001)* & -0.005(0.003) & 0.003(0.001)*** & -0.007(0.002)*** \tabularnewline
\% household with kids & 0.012(0.003)*** & 0.028(0.008)*** & 0.013(0.002)*** & 0.014(0.005)*** \tabularnewline
\% employees working from home & -0.018(0.008)** & 0.010(0.019) & -0.016(0.005)*** & 0.015(0.011) \tabularnewline
Num of occupied nursing home beds per 100 people & 0.036(0.010)*** & 0.086(0.024)*** & 0.008(0.007) & 0.059(0.015)*** \tabularnewline
\% household without health insurance & -0.018(0.011)* & 0.046(0.025)* & -0.003(0.007) & 0.056(0.015)*** \tabularnewline
Insurance $\times$ group effect 1 & 0.062(0.017)*** & 0.088(0.041)*** & 0.046(0.012)*** & 0.001(0.025)* \tabularnewline
Insurance $\times$ group effect 2 & 0.042(0.014)*** & 0.010(0.033) & 0.021(0.010)** & -0.031(0.021) \tabularnewline
Insurance $\times$ group effect 3 & 0.001(0.013) & -0.008(0.031) & 0.018(0.009)** & -0.008(0.019) \tabularnewline
Age group over 65 & 0.014(0.005)*** & 0.069(0.011)*** & 0.008(0.003)*** & 0.043(0.007)*** \tabularnewline
\% Public housing area & -0.005(0.003)* & 0.006(0.007) & -0.002(0.002) & 0.009(0.004)** \tabularnewline

\hline
\multicolumn{4}{m{10.5cm}}{\textit{Note: } Standard errors are in parentheses with regression coefficients. \newline *** p-value $<$ 0.01, ** p-value $<$ 0.05, * p-value $<$ 0.1}

\end{tabular}
\end{table*}

The results of multivariate regression models, which control for neighborhood socioeconomic and demographic covariates, are described in Table \ref{table:multivarite_model}. We combine both ``outflow" groups (``outflow-mixed" and ``outflow-residential") and use the ``stable" neighborhood group as the reference case. After accounting for various demographic characteristics, we continue to observe statistically significant coefficients for the exposure density variables. The positivity rate model (model 3), shows the most dramatic effects of behavior change and measured health incomes. Neighborhoods that reduced exposure density are shown to have 13\% lower positivity rate compared to the reference group. For outflow neighborhoods that maintain the distribution of activities across land use types (classified as the ``outflow-stable" group), the output is a 23\% lower positivity rate. A similar pattern is also found in the case rate model (model 1). %These findings provide some additional empirical evidence to the effectiveness of social distancing as a strategy to reduce COVID-19 spread, emphasizing that proactive neighborhood behavior change can help to prevent transmission of the virus and delay doubling the time of the infection \cite{gao2020mobile,chudik2020voluntary,jay2020neighborhood,matrajt2020evaluating,farboodi2020internal,hatchett2007public,abouk2020immediate}. 

Additionally, race and ethnicity, age group, and socioeconomic status are also found to have statistically significant effects on neighborhood infection rates and outcomes. Communities with larger proportions of minority and lower income populations are more likely to be at risk for virus transmission. For example, for every 10\% increase of Hispanic residents in the community, the positivity rate increases by 5\%, the case rate increases by 9\%, and the death rate increases by 6\%. This finding holds after accounting for changes in exposure density. As expected, exposure density is not shown to be a statistically significant feature in the death rate model (model 2), while the variables related to the presence of vulnerable populations have significant negative impact on the survival probability. We find that elderly population, lack of health insurance coverage, and a high proportion of people living in public housing have positive and statistically significant associations with death rates across the city. Thus, the mortality risk of the virus in socially vulnerable neighborhoods are higher than in other communities, resulting from pre-existing health conditions and the lack of an adequate healthcare access. This also explains, in part, why the ``outflow-stable" group, with the highest proportion of lower income residents without health insurance, experienced an approximately 43\% higher fatality rate compared to the reference group, despite observed lower infection rates.

\section{Discussion and Conclusion}
We present a novel approach to measuring exposure density at high spatial and temporal resolution. By integrating geolocation data and land use classifications, we are able to establish both the extent of activity in a particular area and the nature of that activity across residential, non-residential, and outdoor activities. This approach is scalable to any areal unit of interest: here we utilize a 250m grid and aggregate to the zipcode level to match geography of reported health data. However, it is possible to apply the same methodology to point locations or grids of any size, and then aggregate the units to other common administrative or political boundaries, such as census tracts, counties, and metropolitan areas. We normalize our data to enable comparative studies between regions and to scale the analysis to other cities with similar land use data resources. 

Our findings demonstrate distinct patterns of activity before and after the stay-at-home order across neighborhoods. These neighborhood patterns are clustered into five unique groups, each exhibiting statistically significant variations in socioeconomic and demographic characteristics of residents. In wealthier neighborhoods of Manhattan and Brooklyn, we observe an exodus of residents leaving for other areas around NYC or regions further afield. Presumably, these residents have the means to relocate to second homes or rental homes that provide a greater degree of (perceived) safety from the virus. In addition to the financial ability to make such a move, residents in these neighborhoods also had, in many cases, the option to work remotely, thus reducing the transaction costs of leaving their primary residence. Conversely, we observed clusters of poor, minority neighborhoods that faced greater infection risk. While some residents in neighborhoods in the ``stable" groups did relocate, the large majority stayed in their communities and continued on with their typical (pre-COVID) routines. As a result, we found that the exposure density in these neighborhoods remain relatively constant over the study period, reflecting the need to commute to work and other places of responsibility, especially given that many of those employed worked in occupations deemed essential services. Finally, we find a cluster of neighborhoods that increased their exposure density due to an increased amount of localized activity. These neighborhoods, characterized by lower density, single-family homes in areas further from the Manhattan central business district, are found to have both a greater volume of activity and more activity taking place in non-residential and outdoor areas than normal. The effect of this local activity was an increase, compared to pre-COVID levels, in the probability of coming in contact with others outside of the household or family unit.

The variation in exposure density has a direct and measurable impact on the risk of infection. In neighborhoods where exposure density decreased the most, we find lower rates of infection, positivity rates, and death rate per capita, controlling for other covariates associated with social determinants of health. The communities hardest hit by the virus were in the ``stable-stable" neighborhoods, where residents faced multiple challenges and risk factors. In addition to continuing their normal activity patterns, and thus exposing themselves to greater risk of infection while commuting and in their place of work, these communities have the largest proportion of minorities, among the lowest median incomes, and the lowest rate of health insurance coverage. These compound risks resulted in these vulnerable communities facing the burden of the highest rate of infection, death rate, and positivity rate in the City during the study period.

Our work highlights the importance of understanding neighborhood activity patterns in evaluating the determinants of health outcomes and risk factors for future infection outbreaks. By measuring exposure density at the community scale, we are able to determine the differential behavioral response to social distancing policies based on local risk factors and socioeconomic inequality. Our results expose the significant disparities in health outcomes for racial and ethnic minorities and lower income households. Exposure density provides an additional metric to further explain and understand the disparate impact of COVID-19 on vulnerable communities.

\section*{Acknowledgements}
The authors would like to thank the New York University Center for Urban Science and Progress (NYU CUSP) Research Computing Facility (RCF) for providing and managing database infrastructure and VenPath, Inc. for providing data.

\section*{Funding}
This work was supported, in part, by grants from the National Science Foundation, No. 1653772 and No. 2028687, and from NYU C2SMART, a USDOT Tier 1 University Transportation Center. Any opinions, findings, and conclusions expressed in this paper are those of authors and do not necessarily reflect the views of any supporting institution. All errors remain the authors. 

\section*{Competing interests}
The authors declare that they have no competing interests. 

% \section*{Data and materials availability}
% All data needed to evaluate the conclusions in the paper are present in the paper and/or the Supplementary Materials. Additional data related to this paper may be requested from the authors

\bibliographystyle{elsarticle-num-names}
\bibliography{main.bib}

\begin{thebibliography}{37}
\providecommand{\natexlab}[1]{#1}
\providecommand{\url}[1]{\texttt{#1}}
\providecommand{\urlprefix}{URL }
\expandafter\ifx\csname urlstyle\endcsname\relax
  \providecommand{\doi}[1]{doi:\discretionary{}{}{}#1}\else
  \providecommand{\doi}[1]{doi:\discretionary{}{}{}\begingroup
  \urlstyle{rm}\url{#1}\endgroup}\fi
\providecommand{\bibinfo}[2]{#2}

\bibitem[{Callaway(2020)}]{callaway2020time}
\bibinfo{author}{E.~Callaway}, \bibinfo{title}{Time to use the p-word?
  Coronavirus enter dangerous new phase}, \bibinfo{journal}{Nature}
  \bibinfo{volume}{579}~(\bibinfo{number}{12}) (\bibinfo{year}{2020})
  \bibinfo{pages}{10--1038}.

\bibitem[{Organization et~al.(2020)}]{world2020coronavirus}
\bibinfo{author}{W.~H. Organization}, et~al., \bibinfo{title}{Coronavirus
  disease 2019 (COVID-19): situation report, 72} .

\bibitem[{Van~Bavel et~al.(2020)Van~Bavel, Baicker, Boggio, Capraro, Cichocka,
  Cikara, Crockett, Crum, Douglas, Druckman et~al.}]{van2020using}
\bibinfo{author}{J.~J. Van~Bavel}, \bibinfo{author}{K.~Baicker},
  \bibinfo{author}{P.~S. Boggio}, \bibinfo{author}{V.~Capraro},
  \bibinfo{author}{A.~Cichocka}, \bibinfo{author}{M.~Cikara},
  \bibinfo{author}{M.~J. Crockett}, \bibinfo{author}{A.~J. Crum},
  \bibinfo{author}{K.~M. Douglas}, \bibinfo{author}{J.~N. Druckman}, et~al.,
  \bibinfo{title}{Using social and behavioural science to support COVID-19
  pandemic response}, \bibinfo{journal}{Nature Human Behaviour}
  (\bibinfo{year}{2020}) \bibinfo{pages}{1--12}.

\bibitem[{Sen-Crowe et~al.(2020)Sen-Crowe, McKenney, and
  Elkbuli}]{sen2020social}
\bibinfo{author}{B.~Sen-Crowe}, \bibinfo{author}{M.~McKenney},
  \bibinfo{author}{A.~Elkbuli}, \bibinfo{title}{Social distancing during the
  COVID-19 pandemic: Staying home save lives}, \bibinfo{journal}{The American
  journal of emergency medicine} .

\bibitem[{Courtemanche et~al.(2020)Courtemanche, Garuccio, Le, Pinkston, and
  Yelowitz}]{courtemanche2020did}
\bibinfo{author}{C.~J. Courtemanche}, \bibinfo{author}{J.~Garuccio},
  \bibinfo{author}{A.~Le}, \bibinfo{author}{J.~C. Pinkston},
  \bibinfo{author}{A.~Yelowitz}, \bibinfo{title}{Did Social-Distancing Measures
  in Kentucky Help to Flatten the COVID-19 Curve?} .

\bibitem[{Gao et~al.(2020)Gao, Rao, Kang, Liang, Kruse, Doepfer, Sethi, Reyes,
  Patz, and Yandell}]{gao2020mobile}
\bibinfo{author}{S.~Gao}, \bibinfo{author}{J.~Rao}, \bibinfo{author}{Y.~Kang},
  \bibinfo{author}{Y.~Liang}, \bibinfo{author}{J.~Kruse},
  \bibinfo{author}{D.~Doepfer}, \bibinfo{author}{A.~K. Sethi},
  \bibinfo{author}{J.~F.~M. Reyes}, \bibinfo{author}{J.~Patz},
  \bibinfo{author}{B.~S. Yandell}, \bibinfo{title}{Mobile phone location data
  reveal the effect and geographic variation of social distancing on the spread
  of the COVID-19 epidemic}, \bibinfo{journal}{arXiv preprint arXiv:2004.11430}
  .

\bibitem[{Remuzzi and Remuzzi(2020)}]{remuzzi2020covid}
\bibinfo{author}{A.~Remuzzi}, \bibinfo{author}{G.~Remuzzi},
  \bibinfo{title}{COVID-19 and Italy: what next?}, \bibinfo{journal}{The
  Lancet} .

\bibitem[{Chudik et~al.(2020)Chudik, Pesaran, and
  Rebucci}]{chudik2020voluntary}
\bibinfo{author}{A.~Chudik}, \bibinfo{author}{M.~H. Pesaran},
  \bibinfo{author}{A.~Rebucci}, \bibinfo{title}{Voluntary and mandatory social
  distancing: Evidence on covid-19 exposure rates from chinese provinces and
  selected countries}, \bibinfo{type}{Tech. Rep.},
  \bibinfo{institution}{National Bureau of Economic Research},
  \bibinfo{year}{2020}.

\bibitem[{Patrick et~al.(2020)Patrick, Stanbrook, and
  Laupacis}]{patrick2020social}
\bibinfo{author}{K.~Patrick}, \bibinfo{author}{M.~B. Stanbrook},
  \bibinfo{author}{A.~Laupacis}, \bibinfo{title}{Social distancing to combat
  COVID-19: We are all on the front line}, \bibinfo{year}{2020}.

\bibitem[{Jay et~al.(2020)Jay, Bor, Nsoesie, Lipson, Jones, Galea, and
  Raifman}]{jay2020neighborhood}
\bibinfo{author}{J.~Jay}, \bibinfo{author}{J.~Bor},
  \bibinfo{author}{E.~Nsoesie}, \bibinfo{author}{S.~K. Lipson},
  \bibinfo{author}{D.~K. Jones}, \bibinfo{author}{S.~Galea},
  \bibinfo{author}{J.~Raifman}, \bibinfo{title}{Neighborhood income and
  physical distancing during the COVID-19 pandemic in the US},
  \bibinfo{journal}{medRxiv} .

\bibitem[{Adolph et~al.(2020)Adolph, Amano, Bang-Jensen, Fullman, and
  Wilkerson}]{adolph2020pandemic}
\bibinfo{author}{C.~Adolph}, \bibinfo{author}{K.~Amano},
  \bibinfo{author}{B.~Bang-Jensen}, \bibinfo{author}{N.~Fullman},
  \bibinfo{author}{J.~Wilkerson}, \bibinfo{title}{Pandemic politics: Timing
  state-level social distancing responses to COVID-19},
  \bibinfo{journal}{medRxiv} .

\bibitem[{Chinazzi et~al.(2020)Chinazzi, Davis, Ajelli, Gioannini, Litvinova,
  Merler, y~Piontti, Mu, Rossi, Sun et~al.}]{chinazzi2020effect}
\bibinfo{author}{M.~Chinazzi}, \bibinfo{author}{J.~T. Davis},
  \bibinfo{author}{M.~Ajelli}, \bibinfo{author}{C.~Gioannini},
  \bibinfo{author}{M.~Litvinova}, \bibinfo{author}{S.~Merler},
  \bibinfo{author}{A.~P. y~Piontti}, \bibinfo{author}{K.~Mu},
  \bibinfo{author}{L.~Rossi}, \bibinfo{author}{K.~Sun}, et~al.,
  \bibinfo{title}{The effect of travel restrictions on the spread of the 2019
  novel coronavirus (COVID-19) outbreak}, \bibinfo{journal}{Science}
  \bibinfo{volume}{368}~(\bibinfo{number}{6489}) (\bibinfo{year}{2020})
  \bibinfo{pages}{395--400}.

\bibitem[{Abouk and Heydari(2020)}]{abouk2020immediate}
\bibinfo{author}{R.~Abouk}, \bibinfo{author}{B.~Heydari}, \bibinfo{title}{The
  immediate effect of covid-19 policies on social distancing behavior in the
  united states}, \bibinfo{journal}{Available at SSRN} .

\bibitem[{Hatchett et~al.(2007)Hatchett, Mecher, and
  Lipsitch}]{hatchett2007public}
\bibinfo{author}{R.~J. Hatchett}, \bibinfo{author}{C.~E. Mecher},
  \bibinfo{author}{M.~Lipsitch}, \bibinfo{title}{Public health interventions
  and epidemic intensity during the 1918 influenza pandemic},
  \bibinfo{journal}{Proceedings of the National Academy of Sciences}
  \bibinfo{volume}{104}~(\bibinfo{number}{18}) (\bibinfo{year}{2007})
  \bibinfo{pages}{7582--7587}.

\bibitem[{Farboodi et~al.(2020)Farboodi, Jarosch, and
  Shimer}]{farboodi2020internal}
\bibinfo{author}{M.~Farboodi}, \bibinfo{author}{G.~Jarosch},
  \bibinfo{author}{R.~Shimer}, \bibinfo{title}{Internal and external effects of
  social distancing in a pandemic}, \bibinfo{type}{Tech. Rep.},
  \bibinfo{institution}{National Bureau of Economic Research},
  \bibinfo{year}{2020}.

\bibitem[{Matrajt and Leung(2020)}]{matrajt2020evaluating}
\bibinfo{author}{L.~Matrajt}, \bibinfo{author}{T.~Leung},
  \bibinfo{title}{Evaluating the Effectiveness of Social Distancing
  Interventions to Delay or Flatten the Epidemic Curve of Coronavirus
  Disease.}, \bibinfo{journal}{Emerging infectious diseases}
  \bibinfo{volume}{26}~(\bibinfo{number}{8}).

\bibitem[{Greenstone and Nigam(2020)}]{greenstone2020does}
\bibinfo{author}{M.~Greenstone}, \bibinfo{author}{V.~Nigam},
  \bibinfo{title}{Does social distancing matter?}, \bibinfo{journal}{University
  of Chicago, Becker Friedman Institute for Economics Working Paper}
  ~(\bibinfo{number}{2020-26}).

\bibitem[{Thunstrom et~al.(2020)Thunstrom, Newbold, Finnoff, Ashworth, and
  Shogren}]{thunstrom2020benefits}
\bibinfo{author}{L.~Thunstrom}, \bibinfo{author}{S.~Newbold},
  \bibinfo{author}{D.~Finnoff}, \bibinfo{author}{M.~Ashworth},
  \bibinfo{author}{J.~Shogren}, \bibinfo{title}{The benefits and costs of using
  social distancing to flatten the curve for COVID-19},
  \bibinfo{journal}{forthcoming, Journal of Benefit-Cost Analysis} .

\bibitem[{Wise et~al.(2020)Wise, Zbozinek, Michelini, Hagan
  et~al.}]{wise2020changes}
\bibinfo{author}{T.~Wise}, \bibinfo{author}{T.~D. Zbozinek},
  \bibinfo{author}{G.~Michelini}, \bibinfo{author}{C.~C. Hagan}, et~al.,
  \bibinfo{title}{Changes in risk perception and protective behavior during the
  first week of the COVID-19 pandemic in the United States} .

\bibitem[{Caley et~al.(2008)Caley, Philp, and McCracken}]{caley2008quantifying}
\bibinfo{author}{P.~Caley}, \bibinfo{author}{D.~J. Philp},
  \bibinfo{author}{K.~McCracken}, \bibinfo{title}{Quantifying social distancing
  arising from pandemic influenza}, \bibinfo{journal}{Journal of the Royal
  Society Interface} \bibinfo{volume}{5}~(\bibinfo{number}{23})
  (\bibinfo{year}{2008}) \bibinfo{pages}{631--639}.

\bibitem[{Atchison et~al.(2020)Atchison, Bowman, Vrinten, Redd, Pristera,
  Eaton, and Ward}]{atchison2020perceptions}
\bibinfo{author}{C.~J. Atchison}, \bibinfo{author}{L.~Bowman},
  \bibinfo{author}{C.~Vrinten}, \bibinfo{author}{R.~Redd},
  \bibinfo{author}{P.~Pristera}, \bibinfo{author}{J.~W. Eaton},
  \bibinfo{author}{H.~Ward}, \bibinfo{title}{Perceptions and behavioural
  responses of the general public during the COVID-19 pandemic: A
  cross-sectional survey of UK Adults}, \bibinfo{journal}{medRxiv} .

\bibitem[{Desai(2020)}]{desai2020urban}
\bibinfo{author}{D.~Desai}, \bibinfo{title}{Urban Densities and the Covid-19
  Pandemic: Upending the Sustainability Myth of Global Megacities},
  \bibinfo{journal}{ORF Occasional Paper}
  \bibinfo{volume}{244}~(\bibinfo{number}{4}).

\bibitem[{Viotti(2020)}]{worldbank}
\bibinfo{author}{L.~Viotti}, \bibinfo{title}{COVID {M}obile {D}ata},
  \bibinfo{howpublished}{\url{https://github.com/worldbank/covid-mobile-data}},
  \bibinfo{note}{accessd: 2020-07-08}, \bibinfo{year}{2020}.

\bibitem[{Facebook(2020)}]{facebook}
\bibinfo{author}{Facebook}, \bibinfo{title}{COVID-10 Mobility Data Network},
  \bibinfo{howpublished}{\url{https://visualization.covid19mobility.org/?date=2020-06-14&dates=2020-03-14_2020-06-14&region=WORLD}},
  \bibinfo{note}{accessd: 2020-06-30}, \bibinfo{year}{2020}.

\bibitem[{Kayes et~al.(2020)Kayes, Islam, Watters, Ng, and
  Kayesh}]{kayes2020automated}
\bibinfo{author}{A.~Kayes}, \bibinfo{author}{M.~S. Islam},
  \bibinfo{author}{P.~A. Watters}, \bibinfo{author}{A.~Ng},
  \bibinfo{author}{H.~Kayesh}, \bibinfo{title}{Automated measurement of
  attitudes towards social distancing using social media: A covid-19 case
  study} .

\bibitem[{Xu et~al.(2020)Xu, Dredze, and Broniatowski}]{xu2020twitter}
\bibinfo{author}{P.~Xu}, \bibinfo{author}{M.~Dredze}, \bibinfo{author}{D.~A.
  Broniatowski}, \bibinfo{title}{The twitter social mobility index: Measuring
  social distancing practices from geolocated tweets}, \bibinfo{journal}{arXiv
  preprint arXiv:2004.02397} .

\bibitem[{Das and James(2020)}]{sensor}
\bibinfo{author}{R.~Das}, \bibinfo{author}{P.~James}, \bibinfo{title}{How smart
  city technology can be used to measure social distancing},
  \bibinfo{howpublished}{\url{https://theconversation.com/how-smart-city-technology-can-be-used-to-measure-social-distancing-135139}},
  \bibinfo{note}{accessd: 2020-06-30}, \bibinfo{year}{2020}.

\bibitem[{Bayham et~al.(????)}]{bayhamcolorado}
\bibinfo{author}{J.~Bayham}, et~al., \bibinfo{title}{Colorado Mobility Patterns
  During the COVID-19 Response. 2020}, \bibinfo{journal}{URL: http://www.
  ucdenver.
  edu/academics/colleges/PublicHealth/coronavirus/Documents/Mobility\%
  20Report\_final. pdf} .

\bibitem[{Allcott et~al.(2020)Allcott, Boxell, Conway, Gentzkow, Thaler, and
  Yang}]{allcott2020polarization}
\bibinfo{author}{H.~Allcott}, \bibinfo{author}{L.~Boxell},
  \bibinfo{author}{J.~Conway}, \bibinfo{author}{M.~Gentzkow},
  \bibinfo{author}{M.~Thaler}, \bibinfo{author}{D.~Y. Yang},
  \bibinfo{title}{Polarization and public health: Partisan differences in
  social distancing during the Coronavirus pandemic}, \bibinfo{journal}{NBER
  Working Paper} ~(\bibinfo{number}{w26946}).

\bibitem[{Painter and Qiu(2020)}]{painter2020political}
\bibinfo{author}{M.~Painter}, \bibinfo{author}{T.~Qiu},
  \bibinfo{title}{Political beliefs affect compliance with covid-19 social
  distancing orders}, \bibinfo{journal}{Available at SSRN 3569098} .

\bibitem[{Coven and Gupta(2020)}]{coven2020disparities}
\bibinfo{author}{J.~Coven}, \bibinfo{author}{A.~Gupta},
  \bibinfo{title}{Disparities in mobility responses to covid-19},
  \bibinfo{type}{Tech. Rep.}, \bibinfo{institution}{NYU Stern Working Paper},
  \bibinfo{year}{2020}.

\bibitem[{Google(2020)}]{google}
\bibinfo{author}{Google}, \bibinfo{title}{COVID-19 Community Mobility Reports},
  \bibinfo{howpublished}{\url{https://www.google.com/covid19/mobility/}},
  \bibinfo{note}{accessd: 2020-06-30}, \bibinfo{year}{2020}.

\bibitem[{Wellenius et~al.(2020)Wellenius, Vispute, Espinosa, Fabrikant, Tsai,
  Hennessy, Williams, Gadepalli, Boulange, Pearce
  et~al.}]{wellenius2020impacts}
\bibinfo{author}{G.~A. Wellenius}, \bibinfo{author}{S.~Vispute},
  \bibinfo{author}{V.~Espinosa}, \bibinfo{author}{A.~Fabrikant},
  \bibinfo{author}{T.~C. Tsai}, \bibinfo{author}{J.~Hennessy},
  \bibinfo{author}{B.~Williams}, \bibinfo{author}{K.~Gadepalli},
  \bibinfo{author}{A.~Boulange}, \bibinfo{author}{A.~Pearce}, et~al.,
  \bibinfo{title}{Impacts of state-level policies on social distancing in the
  united states using aggregated mobility data during the covid-19 pandemic},
  \bibinfo{journal}{arXiv preprint arXiv:2004.10172} .

\bibitem[{Aktay et~al.(2020)Aktay, Bavadekar, Cossoul, Davis, Desfontaines,
  Fabrikant, Gabrilovich, Gadepalli, Gipson, Guevara et~al.}]{aktay2020google}
\bibinfo{author}{A.~Aktay}, \bibinfo{author}{S.~Bavadekar},
  \bibinfo{author}{G.~Cossoul}, \bibinfo{author}{J.~Davis},
  \bibinfo{author}{D.~Desfontaines}, \bibinfo{author}{A.~Fabrikant},
  \bibinfo{author}{E.~Gabrilovich}, \bibinfo{author}{K.~Gadepalli},
  \bibinfo{author}{B.~Gipson}, \bibinfo{author}{M.~Guevara}, et~al.,
  \bibinfo{title}{Google COVID-19 community mobility reports: Anonymization
  process description (version 1.0)}, \bibinfo{journal}{arXiv preprint
  arXiv:2004.04145} .

\bibitem[{Apple(2020)}]{apple}
\bibinfo{author}{Apple}, \bibinfo{title}{Mobility Trends Reports},
  \bibinfo{howpublished}{\url{https://www.apple.com/covid19/mobility}},
  \bibinfo{note}{accessd: 2020-06-30}, \bibinfo{year}{2020}.

\bibitem[{UNACAST(2020)}]{unacast}
\bibinfo{author}{UNACAST}, \bibinfo{title}{Social Distancing Scoreboard},
  \bibinfo{howpublished}{\url{https://www.unacast.com/covid19/social-distancing-scoreboard}},
  \bibinfo{note}{accessd: 2020-06-30}, \bibinfo{year}{2020}.

\bibitem[{Hastie et~al.(2009)Hastie, Tibshirani, and
  Friedman}]{hastie2009elements}
\bibinfo{author}{T.~Hastie}, \bibinfo{author}{R.~Tibshirani},
  \bibinfo{author}{J.~Friedman}, \bibinfo{title}{The elements of statistical
  learning}, \bibinfo{year}{2009}.

\end{thebibliography}

\end{document}